%% file: x86isa.tex
\documentclass[copyright,creativecommons]{eptcs}
\providecommand{\event}{ACL2 Workshop, 2017}
\usepackage{xspace}
\usepackage{graphicx}
\usepackage[english]{babel}
\usepackage[utf8]{inputenc}
\usepackage{times}
\usepackage[T1]{fontenc}
\usepackage{lipsum}
\usepackage{ragged2e}
\usepackage{alltt}
\usepackage{rotating}
\usepackage{lscape}
\usepackage{setspace}
\usepackage{array}
\usepackage{verbatim}
\usepackage{fancyvrb}
\usepackage{upquote}
\usepackage[dvipsnames]{xcolor}
\usepackage{tikz}
\usetikzlibrary{arrows}
\usepackage[htt]{hyphenat}
\usepackage{hyperref}

\allowhyphens

\definecolor{fgreendark}{rgb}{0.0, 0.45, 0.13}
\newcommand{\textitg}[1]{\textcolor{fgreendark}{\textit{#1}}} 

\begin{document}
\title{The {\tt x86isa} Books: Features, Usage, and Future Plans}

\author{Shilpi Goel
  \institute{Department of Computer Science, University of Texas at
    Austin\thanks{The author is now at Centaur Technology, Inc., but
      this work was done as a part of the author's PhD at UT
      Austin.}}
    \email{shigoel@cs.utexas.edu}}

\def\titlerunning{The {\tt x86isa} Books}
\def\authorrunning{S. Goel}

\maketitle

\begin{abstract}
  The {\tt x86isa} library, incorporated in the ACL2 community books
  project, provides a formal model of the x86 instruction-set
  architecture and supports reasoning about x86 machine-code programs.
  However, analyzing x86 programs can be daunting --- even for those
  familiar with program verification, in part due to the complexity of
  the x86 ISA.  Furthermore, the {\tt x86isa} library is a large
  framework, and using and/or contributing to it may not seem
  straightforward.  We present some typical ways of working with the
  {\tt x86isa} library, and describe some of its salient features that
  can make the analysis of x86 machine-code programs less arduous.  We
  also discuss some capabilities that are currently missing from these
  books --- we hope that this will encourage the community to get
  involved in this project.
\end{abstract}




\input{intro}

\input{overview}

\input{concrete-program-run}

\input{code-proofs}

\input{future-plans}

\input{conclusion}

\section*{Acknowledgements}
This work was supported by DARPA under contract number
N66001-10-2-4087, and by NSF under contract number CNS-1525472.  I am
grateful to Warren Hunt, Jr., Matt Kaufmann, J Moore, and Robert
Watson for their guidance.  I thank Eric Smith for helpful discussions
about his use of and expectations from the {\tt x86isa} books, and
Anna Slobodova for feedback on an early version of this paper.

\vspace{-2mm}
\bibliographystyle{eptcs}
\bibliography{x86isa}

\end{document}

%% file: intro.tex
\section{Introduction}\label{sec:introduction}
The ACL2 community books contain several machine models
(\href{https://github.com/acl2/acl2/tree/master/books/models/y86}{\underline{Y86}},
\href{https://github.com/acl2/acl2/tree/master/books/models/jvm}{\underline{JVM}},
etc.) and libraries that aid in program verification
(\href{https://github.com/acl2/acl2/tree/master/books/coi}{\underline{COI}},
\href{https://github.com/acl2/acl2/tree/master/books/projects/stateman}{\underline{Stateman}},
\href{https://github.com/acl2/acl2/tree/master/books/projects/codewalker}{\underline{Codewalker}},
etc.).  The {\tt {x86isa}} library
(\href{http://www.cs.utexas.edu/users/moore/acl2/manuals/current/manual/?topic=ACL2____X86ISA}{{\tt
    :doc \underline{x86isa}}}) adds to this repertoire by providing
yet another formal, executable machine model --- that of the
single-processor x86 instruction-set architecture, with a
specification of 400+ opcodes executing in Intel's 64-bit mode of
operation. This library also offers the following capabilities:
\begin{itemize}
\item A tool to read, parse, and load an executable file
  (Mach-O~\cite{mach-o} and ELF~\cite{systemvabi} formats) at the
  appropriate memory location of the x86 state;
\item Utilities along the lines of the GNU Debugger (GDB) and
  Pintool~\cite{pintool} to monitor concrete program runs, and
\item Books that provide rules that facilitate symbolic simulation of
  x86-64 machine-code programs.
\end{itemize}
Also, there are examples that illustrate how the above were used to
set up the model for a program run, dynamically instrument a program,
run co-simulations against an actual x86 processor for model
validation, and perform x86 machine-code proofs.  Consequently, the
{\tt x86isa} library, which is still in active development, is large
--- currently, it consists of around 60,000 lines of ACL2 (not
counting automatically generated events) and around 240 files.

Just the complexity and size of the x86 ISA can deter people from
being serious practitioners of x86 machine-code verification.
Therefore, formal tools built to support this undertaking have an
obligation to be easily accessible to the users --- at least to those
who already have some familiarity with program verification.  To this
end, we describe the {\tt x86isa} library so that a user can find it
relatively straightforward to get started with x86 machine-code
analysis.  We also present some important capabilities that are
currently missing from this library.  We hope that this will encourage
the ACL2 community to contribute to this project, both by way of
adding new features themselves and by way of providing feedback that
will help {\tt x86isa} become sophisticated over time.

%% file: overview.tex
\section{Overview}\label{sec:overview}
The x86 ISA model has been developed using the classical {\em
  interpreter-style of operational semantics} that is prevalent in the
ACL2 community: a recursively-defined interpreter over the x86 state
is used to ascribe semantics to x86-64 programs.  Models written using
this style have the following main components: a machine state,
semantic functions that describe instructions' behavior, a step
function that executes one instruction by calling the appropriate
instruction semantic function, and a run function (i.e., the
interpreter) that calls the step function iteratively.  We briefly
describe our x86 ISA model in this section, referring an interested
reader elsewhere~\cite{x86isa-procos,goel-dissertation} for details.

\subsection{x86 State}\label{sec:state}
The x86 ISA state has been specified using an abstract
stobj~\cite{abstract-stobjs} called {\tt x86}. The x86 ISA components
currently supported by our model are: general-purpose registers ({\tt
  rax}, {\tt rbx}, etc.), instruction pointer, flags register, segment
registers ({\tt cs}, {\tt ss}, etc.), memory-management registers
({\tt gdtr}, {\tt ldtr}), interrupt/task management registers ({\tt
  idtr}, {\tt tr}), control registers ({\tt cr0}, {\tt cr1}, etc.),
floating-point registers (e.g., {\tt fp-data0}, {\tt mmx0}\footnote{As
  dictated by the x86 ISA, MMX registers are aliased to the low 64
  bits of the FPU's data registers.}, {\tt fp-ctrl}, {\tt fp-status},
etc.), XMM registers, MXCSR register, machine-specific
registers\footnote{Intel defines many MSRs. Our model currently
  supports only 6 of them: {\tt ia32\_efer}, {\tt ia32\_fs\_base},
  {\tt ia32\_gs\_base}, {\tt ia32\_kernel\_gs\_base}, {\tt
    ia32\_lstar}, {\tt ia32\_star}, {\tt ia32\_fmask}.}, and a
byte-addressable main memory that specifies 2$^{52}$ bytes.
Additionally, {\tt x86} contains some fields that control and report
on the model's operation, rather than that of the machine.  An example
of such a field is the model state {\tt ms} --- if a model-related
error occurs at any point during the execution of a program (e.g., an
unimplemented opcode is encountered), then this field is populated
with information about the error and execution is halted.  Thus, the
x86 ISA model is expected to reflect the real machine's state only if
the {\tt ms} field is empty.  We discuss other such fields --- the
{\tt user-level-mode}, {\tt page-structure-marking-mode}, {\tt undef},
{\tt os-info}, and {\tt env} --- later in this paper.

\subsection{Modes of Operation and x86 Memory Interface}\label{sec:modes-of-operation}
Reasoning about {\em all} of the x86 machine code involved in the
execution of a user-level (application) program is a huge undertaking.
In addition to the x86 code corresponding to the application program
itself, one would need to consider the x86 code corresponding to the
underlying system programs as well.  For example, consider a C program
that uses {\tt printf} to print ``Hello, world!'' to standard output
--- {\tt printf} is a standard C library function that ultimately
relies on the {\tt write} system call provided by the OS.  Statically
compiling this program on an x86 platform generates an executable file
of size 0.8MB!\footnote{This program was statically compiled on an
  Intel Xeon CPU (E31280) using the default options of GCC compiler,
  version 4.8.4.  The standard C library used was Ubuntu EGLIBC,
  version 2.19.}

For expediency during application program verification, a user may
wish to assume, either temporarily or permanently, that the underlying
system programs behave as expected.  To this end, the x86 ISA model
provides two main modes of operation: the {\em system-level mode} and
the {\em user-level mode}.  The x86 ISA model operates in the
user-level mode when the {\tt user-level-mode} field in {\tt x86} is
non-nil; otherwise, it operates in the system mode.  Furthermore, the
system-level mode offers two sub-modes of operation --- the {\em
  marking} and {\em non-marking} mode; a non-nil value in the field
{\tt page-structure-marking-mode} dictates that the model operate in
the system-level marking mode.  These two sub-modes are used to
optimize reasoning about certain kinds of system programs, and are
discussed later in Section~\ref{sec:system-prog-verification}.  For
now, we just note that the ``true'' specification of the x86 ISA is
given by the model's system-level marking mode of operation, and any
discussion about the system-level mode pertains to this sub-mode
unless specified otherwise.

The system-level mode is intended to provide the same environment to a
program as is provided by an x86 processor, and is suitable for the
verification of OS routines.  The user-level mode is intended for the
verification of application programs under the assumption that the
relevant OS services are correct.  In this mode, the x86 system state
--- which includes some memory-resident data structures like the page
tables --- is hidden from the programs.  The system-level and
user-level modes share a large part of their code base, but they
differ significantly in the view of their memory and the
implementation of certain instructions --- we discuss the latter in
Section~\ref{sec:semantic-functions}.

The x86 processors offer two main kinds of memories --- {\em linear
  memory} and {\em physical memory} --- which are indexed by linear
and physical addresses, respectively.  Physical memory is the RAM
addressed by a processor on its bus.  Linear memory is an abstraction
of the physical memory that is offered to x86 programs via a memory
management mechanism called {\em paging}.  Paging is used to map a
linear address to physical address using information present in
ISA-specified, memory-resident data structures called the {\em page
  tables}.  64-bit programs cannot access physical memory directly;
however, privileged 64-bit programs can alter the linear-to-physical
address mapping by modifying the page tables.

The system-level mode of the x86 ISA model includes a specification of
paging, and thus, it has a model of both linear and physical memory.
In this mode, every linear memory access is translated to the
corresponding physical memory access.  The user-level mode has a model
only of the linear memory because application programs typically do
not get adequate privileges for directly interacting with the system
data structures.  The same memory field in {\tt x86} is configured to
specify physical memory in the system-level mode and linear memory in
the user-level mode.  To facilitate code sharing between these modes,
we provide a uniform linear memory interface, where top-level memory
accessor and updater functions call the appropriate mode-specific
functions.  This prevents us from needing to define two versions of an
x86 ISA specification function.

Both the modes of operation specify yet another x86 memory management
mechanism: {\em segmentation}.  The system-level mode models
segmentation in its full detail, whereas the user-level model captures
only its application-level view.  We omit details about this mechanism
here because segmentation is mostly disabled in the 64-bit mode.

\subsection{Instruction Semantic Functions}\label{sec:semantic-functions}
The behavior of each instruction can be defined in terms of reads from
and writes to the x86 state.  For example, an {\tt add} instruction
reads the source operands from the x86 state and then writes the
following to the x86 state: the appropriately-sized sum, the updated
flags, and the new value of the instruction pointer.  Of course, this
is a largely incomplete description of {\tt add}; we have omitted
important details --- such as how operand sizes are determined, when
exceptions are thrown, etc. --- from this description.

We have modeled 413 x86 instruction opcodes, including arithmetic,
floating-point, and control-flow instructions.  The x86 ISA model also
contains a specification of some system-mode instructions like {\tt
  lgdt}, {\tt lldt}, {\tt lidt}, etc. --- these instructions are
available only in the system-level mode of operation. The list of
instructions that are specified by our x86 ISA model can be found at
\href{http://www.cs.utexas.edu/users/moore/acl2/manuals/current/manual/?topic=X86ISA____IMPLEMENTED-OPCODES}{{\tt
    :doc \underline{x86isa::implemented-opcodes}}}.

\subsubsection{Undefined and Random Values}\label{sec:undef-and-random}
An important part of defining instruction semantic functions is
accounting for undefined and/or random behavior that is inherent in
certain instructions.  For example, many commonly-used instructions
like {\tt mul} and {\tt div} leave certain flags undefined, and the
{\tt rdrand} instruction returns random values.  We specify undefined
and random values with the function {\tt undef-read}, which simply
invokes the function {\tt undef-read-logic}.
\begin{Verbatim}[commandchars=\\\{\}]
(defun undef-read-logic (x86)
  \textitg{ ;; Declarations elided.}
  (let* ((undef-seed (nfix (undef x86)))
         (new-unknown (create-undef undef-seed))
         (x86 (!undef (1+ undef-seed) x86)))
    (mv new-unknown x86)))

(defun-notinline undef-read (x86)
  \textitg{;; Declarations elided.}
  (undef-read-logic x86))
\end{Verbatim}
The functions {\tt undef} and {\tt !undef} are the native accessor and
updater functions of the {\tt undef} field in the x86 state.  The
function {\tt create-undef} is a constrained function, and its only
known property is that it always returns a {\tt natp}.  After
admitting {\tt undef-read-logic}, we make {\tt !undef} untouchable
(see
\href{http://www.cs.utexas.edu/users/moore/acl2/manuals/current/manual/index.html?topic=ACL2____PUSH-UNTOUCHABLE}{{\tt
    :doc \underline{push-untouchable}}}) to ensure that {\tt
  undef-read-logic} is the only function that can modify this {\tt
  undef} field.  Also, we never use {\tt create-undef} in any function
other than {\tt undef-read-logic}.

The upshot of all of this arrangement is that {\tt undef-read-logic}
always returns an {\em indeterminate} value that can be used to
specify either an undefined or a random value\footnote{Indeterminate
  values have the following useful property: the result of an equality
  test of an indeterminate value with any other value is unknown.}.
Every call of {\tt undef-read-logic} produces a value that is equal to
{\tt create-undef} invoked with the current value of the {\tt undef}
field, and the {\tt undef} field is incremented every time {\tt
  undef-read-logic} is called.  Since {\tt !undef} and {\tt
  create-undef} are never used outside {\tt undef-read-logic}, {\tt
  create-undef} always gets unique arguments.  Essentially, this
arrangement gives us a pool of indeterminate values that can be used
when required.

We need to model all possible behaviors resulting from an
indeterminate value while reasoning, but an appropriate concrete value
is needed during execution.  To this end, we use {\tt undef-read}
(instead of {\tt undef-read-logic}) as our top-level specification
function, and re-define it under the hood using
\href{http://www.cs.utexas.edu/users/moore/acl2/manuals/current/manual/?topic=ACL2____INCLUDE-RAW}{{\tt
    \underline{include-raw}}} so that {\tt undef-read-logic} is used
for reasoning and {\tt undef-read-exec} is used for execution; we omit
the definition of {\tt undef-read-exec} here.  Note that {\tt
  undef-read} is directed to not be inlined by the Lisp compiler
because re-definition of inlined functions may result in unpredictable
behavior.
\begin{Verbatim}[commandchars=\\\{\}]
(defun undef-read$notinline (x86)
  \textitg{;; Declarations elided.}
  (when
      \textitg{;; When the x86 model is being used for reasoning:}
      (or (equal (f-get-global
                  'in-prove-read ACL2::*the-live-state*)
                 t)
          (equal (f-get-global
                  'in-verify-read ACL2::*the-live-state*)
                 t))
    (return-from X86ISA::undef-read$notinline
      (X86ISA::undef-read-logic x86)))
  \textitg{;; When the x86 model is being used for concrete execution:}
  (undef-read-exec x86))
\end{Verbatim}

\subsubsection{System Calls and Non-Determinism}\label{sec:syscalls-non-det}
In addition to the memory model and availability of system-level
instructions, a significant difference between the system-level and
user-level modes is in their treatment of system calls.  System calls
are requests for services made by an application program to the OS.
The {\tt syscall} instruction, used by application programs, transfers
control to a more privileged system sub-routine.  The corresponding
instruction {\tt sysret} is used by system sub-routines to transfer
control back to the application program.  In the system-level mode,
these two instructions are modeled as per their specifications in the
Intel ISA manuals~\cite{intel-manuals}.  In the user-level mode, the
specification of {\tt syscall} has been extended to provide the
semantics of some commonly used system calls like {\tt read}, {\tt
  write}, {\tt open}, {\tt close}, {\tt lseek}, {\tt dup}, {\tt link},
and {\tt unlink}.  These system calls are OS-specific --- for
instance, the specification of {\tt read} on FreeBSD is somewhat
different from that on Linux.  The {\tt os-info} field in the x86
state is used to identify the OS under consideration so that the
appropriate semantic function for these system calls can be chosen.
The extended semantics of {\tt syscall} is intended to capture system
call behavior in its entirety, right from its invocation to its
return, and therefore, the {\tt SYSRET} instruction is unavailable in
the user-level mode.  We validate our system call specification
functions by running co-simulations against the real machine plus the
chosen OS.

System calls can exhibit different behaviors for different runs, even
if given the same inputs --- thus, they are non-deterministic from the
point of view of an application programmer.  Consider a {\tt read}
system call that is invoked to read from a file.  It is possible that
one run be successful, but another result in failure if the file has
been deleted.  In order to formally characterize the interaction of an
application program with the underlying OS, we model an external
environment using the {\tt env} field in {\tt x86} --- this field
represents the part of the external world that affects or is affected
by system calls.  The {\tt env} field includes a file system and an
oracle sub-field that specifies the result of non-deterministic
computations.  For example, the file descriptor (or handle) assigned
to a file by the {\tt open} system call is the lowest-numbered 32-bit
file descriptor not currently open for that process --- this
descriptor may be different for different invocations of the system
call, and thus, we obtain it from the oracle.  The oracle is a map of
linear addresses to a list of values; if it needs to be consulted
during the execution of an instruction, then the first value in the
list corresponding to the address of the instruction is returned.  It
is the user's responsibility to initialize {\tt env}, and hence, the
oracle, appropriately while reasoning --- this provides a way to state
precisely the expectations from the environment.  For instance, when
reasoning about the {\tt open} system call, the {\tt env} field in the
initial x86 state can be constrained in such a way that the oracle
returns an arbitrary 32-bit natural number that can be used as a file
descriptor.

Our system call specification functions are re-defined in the same
manner as {\tt undef-read} so that foreign C/Assembly
functions\footnote{We rely on CCL's
  \href{http://ccl.clozure.com/manual/chapter13.html}{\underline{Foreign
      Function Interface}} for the execution of system calls.} are
invoked during concrete executions --- these foreign functions request
the system call service from the underlying OS (i.e., the host OS
running ACL2) and return the results to the ACL2 caller function.  The
{\tt os-info} field (and for that matter, {\tt env} too) is irrelevant
during concrete executions.

Note that the {\tt env} field could have been used to specify
undefined and random values too --- the user could constrain the
oracle to contain appropriate symbolic values (corresponding to the
undefined and random values) at appropriate linear addresses
(corresponding to the linear addresses of the instructions that
generate these indeterminate values).  Why then do we use the
arrangement with the {\tt undef} field?  One reason is that these
fields serve separate purposes.  The {\tt env} field is used to
specify non-deterministic behavior resulting from reliance on an
external environment, whereas the {\tt undef} field is used to model
indeterminateness in the ISA itself.  Another reason is convenience
--- if {\tt env} were used instead of {\tt undef}, the user would have
to initialize the oracle in {\tt env} whenever instructions that write
undefined or random values in some state components are to be
executed.  Such instructions --- especially those that leave some
flags undefined, like {\tt mul} and {\tt div} --- are encountered
frequently in a typical program, and using the {\tt undef} field saves
the user quite a lot of work.

\subsection{Interpreter}\label{sec:interpreter}
The x86 step function, called {\tt x86-fetch-decode-execute}, fetches
the next instruction from the memory (which is located at the address
in the instruction pointer register {\tt rip}), decodes it, and then
calls the appropriate instruction semantic function.  The run
function, {\tt x86-run}, is the x86 ISA interpreter.  Its definition
is straightforward:
\begin{Verbatim}[baselinestretch=1,commandchars=\\\{\}]
(defun x86-run (n x86)
  \textitg{;; Declarations elided.}
  \textitg{;; Halt if there is a problem indicated by the ms}
  \textitg{;; field, or if there are no more instructions}
  \textitg{;; left to execute.}
  (cond ((ms x86) x86)
        ((zp n) x86)
        (t (let* ((x86 (x86-fetch-decode-execute x86))
                  (n (1- n)))
             (x86-run n x86)))))
\end{Verbatim}

%% file: concrete-program-run.tex
\section{Dynamic Instrumentation of x86 Programs}\label{sec:concrete-program-run}
Like most machine models written in ACL2, the x86 ISA model is also
executable.  The execution speed of the model is around 3.3 million
instructions/second in the user-level mode and around 320,000
instructions/second in the system-level mode with a set-up of 1G
page-table configuration\footnote{This speed was measured on a Intel
  Xeon E31280 CPU @ 3.50GHz with 32GB RAM.}.  One can validate the
model against a real x86 processor by performing co-simulations.  The
model can be used as an instruction-set simulator to inspect the
behavior of x86 machine-code programs by running concrete tests.  It
is generally a good idea to run such tests before reasoning about a
program --- testing may reveal ``obvious'' bugs quickly and may also
help in program comprehension.  We describe how to set up the x86 ISA
model for a concrete run of a program and how to dynamically
instrument a program run.

\subsection{Initializing the x86 ISA Model for a Concrete Run}\label{sec:x86-initialize}
If not already available, obtain the x86 machine-code version of the
program to be executed --- for instance, a given C source program {\tt
  foo.c} can be compiled on an x86 machine using GCC or LLVM to obtain
the executable file {\tt foo.o}.  The file {\tt foo.o} contains
information that is necessary to execute the program, such as the x86
machine code itself, program's data, linear addresses where the
various sections of the program must be placed, etc.  Independently of
the x86 ISA model, one can examine executable files using tools such
as \href{http://man7.org/linux/man-pages/man1/objdump.1.html}{{\tt
    \underline{objdump}}} and
\href{https://www.freebsd.org/cgi/man.cgi?query=otool&sektion=1&manpath=Darwin+8.0.1\%2fppc}{{\tt
    \underline{otool}}}.

Now all we have to do is arrange for {\tt foo.o} to be read in and
parsed by the x86 model and then initialize the x86 state
appropriately based on the information in {\tt foo.o}.  To this end,
we recommend creating a fresh file, say {\tt foo-run.lsp}, which
contains the following events:
\begin{enumerate}

\item Include the top-level {\tt x86isa} book.
  \begin{Verbatim}
    (in-package "X86ISA")
    (include-book "projects/x86isa/top" :dir :system :ttags :all)
  \end{Verbatim}

\item The default value of {\tt user-level-mode} field in the x86
  state is {\tt t}.  Thus, if operation in the user-level mode is
  required (probably because {\tt foo.c} is an application program),
  then go to the next step.  However, in the system-level mode, the
  x86 state includes a model of the physical memory.  Since {\tt
    foo.o} contains memory locations in the linear address space, the
  paging data structures must be set up {\em before} {\tt foo.o} is
  read and loaded into the x86 state.

  We provide a function {\tt init-system-level-mode} that switches the
  model to the system-level mode by writing {\tt nil} to the {\tt
    user-level-mode} field and loads our default configuration of
  paging data structures in the model's physical memory at a specified
  address, say $0$ for this contrived example.  This value $0$ is also
  written to the control register {\tt cr3} so that the processor
  knows where to locate the paging data structures in the physical
  memory.  Our default data structures simply provide an identity
  mapping from linear to physical addresses.
  \begin{Verbatim}[commandchars=\\\{\}]
    (init-system-level-mode 0 x86)
  \end{Verbatim}
  A user can choose to load his own configuration of paging data
  structures by writing to the physical memory in the x86 state
  directly.

\item The program in {\tt foo.o} can be read and loaded into the
  memory of the x86 state by using {\tt binary-file-load}.  At this
  time, {\tt binary-file-load} supports only ELF~\cite{systemvabi}
  (commonly used on Unix systems) or Mach-O~\cite{mach-o} (commonly
  used on FreeBSD/Darwin systems) binaries.
  \begin{Verbatim}[commandchars=\\\{\}]
    (binary-file-load "foo.o")
  \end{Verbatim}
  Note that if instead of an executable file, the x86 machine-code
  program is available simply as a list of bytes intended to be
  located at a particular linear memory location (or some other such
  formulation), it is straightforward to load it into the model by
  simply writing these bytes to the memory --- see the following step.

\item Other components of the x86 state, like the instruction pointer,
  registers, etc., can be initialized by using {\tt init-x86-state}.
  \begin{Verbatim}[commandchars=\\\{\}]
    (init-x86-state
      \textit{<initial contents of MS field --- typically} nil\textit{>}
      \textit{<initial value of the instruction pointer>}
      \textit{<linear address where execution should halt>}
      \textit{<initial values of various registers...>}
      \textit{<updates to the memory>}
     x86)
   \end{Verbatim}
  Alternatively, one may use the stobj's native updater functions to
  write to the x86 state.

 \item Run the program by executing {\tt x86-run} --- the run function
   will either execute {\em <n>} number of steps, or terminate early
   if either an error is encountered or if the instruction pointer
   contains the linear address where execution is instructed to halt
   (see third argument of {\tt init-x86-state} above).
   \begin{Verbatim}[commandchars=\\\{\}]
     (x86-run \textit{<n>} x86)
   \end{Verbatim}
   The contents of the {\tt ms} field after the termination of this
   run will indicate whether the program ran successfully or not.  One
   can also dynamically debug the program, as described in
   Section~\ref{sec:x86-monitor-run} below.  Upon the successful
   completion of the program, its output can be inspected by reading
   the relevant components in the final x86 state.
\end{enumerate}

Note that if one needs to perform another concrete run of the program
in the same ACL2 session, the x86 state must be initialized again ---
in particular, the instruction pointer must point to the first
instruction to be executed and the {\tt ms} field must be {\tt nil}.

All these utilities aside, how does one determine the initial values
of the components of the x86 state?  A user familiar with x86 assembly
and/or machine code may be able to figure this out simply by
``reading'' the program --- this task may be easier if the high-level
source program is available too.  A possibly less time-consuming
alternative is to run the program on the real machine (or an
instruction-set simulator like QEMU~\cite{qemu}) and take a snapshot
of the contents of registers, flags, memory locations,
etc. immediately before the first instruction to be executed.  The x86
model's state can be initialized with all of these values.  This
second approach has the benefit that the model's initial state matches
that of the real machine, which helps in model validation via
co-simulations.

The topic
\href{http://www.cs.utexas.edu/users/moore/acl2/manuals/current/manual/?topic=X86ISA____PROGRAM-EXECUTION}{{\tt
    :doc \underline{x86isa::program-execution}}} contains more
information on initializing the x86 state.  Examples of such {\tt
  foo-run.lsp} files for various programs are also available
\href{https://github.com/acl2/acl2/tree/master/books/projects/x86isa/tools/execution/examples}{\underline{online}}.

\subsection{Monitoring a Concrete Run on the x86 ISA Model}\label{sec:x86-monitor-run}
Tools like the GDB (GNU Debugger) and Pintool are used to monitor x86
machine-code programs at runtime.  GDB allows profiling by executing a
program one instruction at a time, inserting breakpoints, etc.,
whereas Pintool injects instrumentation code into the program
itself\footnote{It should be noted that instrumentation code, such as
  that included by Pintool, is supposed to be transparent to the
  program; however, it is not unexpected for such code to
  inadvertently alter the behavior of the program.  Instead of
  injecting x86 machine code, our utilities monitor the ACL2
  specification functions of our x86 model.}.  We provide some
utilities in the {\tt x86isa} library that mimic these tools; the
capabilities currently provided are as follows:
\begin{itemize}
\item Stepping the interpreter once, \`a la {\tt stepi} command of GDB;
\item Stepping the interpreter {\tt n} instructions at a time;
\item Inserting breakpoints where the execution of the program will
  stop: Arbitrary ACL2 functions can be used to define these stopping
  points.  An illustrative example is as follows: one can write an
  ACL2 function that computes the sum of values in a range of memory
  addresses and insert a breakpoint that instructs the x86 interpreter
  to halt as soon as the value returned by this function becomes
  greater than the current value of the {\tt rax} register.
\item Logging all memory read and write operations;
\item Logging the x86 state (sans the memory) --- either the current
  x86 state or the x86 state obtained after every instruction or after
  every breakpoint can be logged to a file.
\end{itemize}
The ability to log the x86 state is useful in co-simulations --- one
can simply compare the logs generated by the program to those by the
model in order to perform model validation.  Syntax and usage of our
monitoring utilities are described at
\href{http://www.cs.utexas.edu/users/moore/acl2/manuals/current/manual/?topic=X86ISA____DYNAMIC-INSTRUMENTATION}{{\tt
    :doc \underline{x86isa::dynamic-instrumentation}}}.

%% file: code-proofs.tex
\section{Formal Analysis of x86 Programs}\label{sec:code-proofs}
Given an ACL2 machine model defined using operational semantics,
various code proof styles can be used for program verification.  We do
not discuss them in this paper, and refer the reader
elsewhere~\cite{raymoore} for details.  However, central to almost all
these proof styles is the ability to symbolically simulate a program.
The {\tt x86isa} books provide the usual ACL2 rules that enable
symbolic simulation of x86 machine-code programs by controlling the
{\em unwinding} of the interpreter.

\begin{enumerate}
\item {\bf Step Function Opener Rule:} This rule dictates the
  conditions under which a call of the step function, {\tt
    x86-fetch-decode-execute}, should be expanded by ACL2.  For
  instance, one of these conditions is that the {\tt ms} field in the
  initial x86 state should be {\tt nil}.  Because of this rule, ACL2
  first expands that call of the step function about which enough
  information to resolve the hypotheses of this rule is known.
  Typically, this means expanding the call corresponding to the
  current instruction (i.e., the instruction located at the address
  contained in the instruction pointer).
\item {\bf Run Function Sequential Composition Rule:} This rule
  facilitates compositional reasoning by reducing the problem of
  reasoning about {\tt (n1 + n2)} number of instructions to two
  smaller problems --- first reasoning about {\tt n1} instructions,
  and then about {\tt n2} instructions.  That is, it rewrites
  expressions of the form {\tt (x86-run (clk+ n1 n2) x86)} to {\tt
    (x86-run n2 (x86-run n1 x86))} when applicable.
\item {\bf Run Function Opener Rule:} This rule controls the expansion
  of the run function by rewriting {\tt (x86-run n x86)} to {\tt
    (x86-run (1- n) (x86-fetch-decode-execute x86))}, provided that
  the {\tt ms} field is {\tt nil} and {\tt n} is not equal to zero.
\end{enumerate}

Additionally, the {\tt x86isa} books also contain read-over-write and
write-over-write rules that describe the interaction between the x86
state accessor and updater functions using the notions of
non-interference and overlap.  An example of a simple non-interference
property is that a write to a register {\tt i} does not interfere with
a subsequent read from a register {\tt j}, provided that {\tt i} and
{\tt j} are distinct.  Analogously, an example of a simple overlap
property is that if consecutive writes are made to register {\tt i},
then the most recent write will be the only visible one.

Thus, the {\tt x86isa} books include the typical ACL2 rules that will
be immediately familiar to a user with some experience in code proofs.
These books provide lemma libraries to support almost completely
automated symbolic simulation of many x86 programs --- we have also
documented how a user can debug a failed attempt at unwinding the x86
interpreter
(\href{http://www.cs.utexas.edu/users/moore/acl2/manuals/current/manual/?topic=X86ISA____DEBUGGING-CODE-PROOFS}{{\tt
    :doc \underline{x86isa::debugging-code-proofs}}}).  We now give
some examples of reasoning about x86 machine-code programs as a way to
illustrate the different methodologies a user can adopt when working
with the {\tt x86isa} books.

\subsection{Verifying Application Programs}\label{sec:user-prog-verification}
The user-level mode of operation of the x86 ISA model is well-suited
to application program verification, due to reasons discussed
previously in Section~\ref{sec:modes-of-operation}.  We consider the
following two kinds of application programs here: the first is
structurally simple and contains straight-line code that performs
dense arithmetic and logical operations on fixed-width inputs (e.g.,
sub-routines that do bit twiddling), and the second contains loops,
branches, and maybe even makes some system calls (e.g., a sub-routine
that computes the word-count of a file).

Using the {\tt x86isa} books, one can choose to verify the first
program completely automatically --- without using any rules provided
by the {\tt x86isa} books --- by using the bit-blasting capabilities
of
\href{http://www.cs.utexas.edu/users/moore/acl2/manuals/current/manual/?topic=ACL2____GL}{
  \underline{GL}}~\cite{gl-diss,bit-blasting-GL} but at the expense of
a general theorem of correctness.  That is, one may need to constrain
the initial x86 state by assigning concrete values instead of symbolic
ones to certain components in order to reduce the complexity of the
AIGs generated by GL, thereby making the problem tractable for
bit-blasting.  An example of such a theorem is {\tt
  x86-popcount-64-correct} below, which states that a given program
that is intended to compute the population count of its 64-bit input
{\tt n} meets its specification
(\href{https://github.com/acl2/acl2/tree/master/books/projects/x86isa/proofs/popcount/popcount.lisp}{\underline{proof
    script}} and a detailed description~\cite{goel-vstte-13} are
available).  This theorem is in terms of the program being located at
fixed addresses instead of being (mostly)
position-independent\footnote{We say ``mostly'' because a program's
  location, even a parameterized one, needs to be constrained in some
  important ways so that it does not overlap with the stack, data,
  etc.} --- note that {\tt *popcount-64*} is a list of pairs of fixed
linear addresses and the program's bytes.
\begin{Verbatim}[commandchars=\\\{\}]
(defconst *popcount-64*
  (list
   (cons #x400650 #x89) \textitg{;; mov %edi,%edx}
   (cons #x400651 #xfa)
   \textitg{;; ...               ... many instructions elided ...}
   (cons #x4006c2 #xc3) \textitg{;; retq}
   ))

(def-gl-thm x86-popcount-64-correct
  :hyp (and (natp n)
            (< n (expt 2 64)))
  :concl
  (b* ((start-address #x400650)
       (halt-address #x4006c2)
       \textitg{;; Assigning default values to state components:}
       (x86 (create-x86))
       (x86 (!user-level-mode t x86))
       ((mv flg x86)
        (init-x86-state
         nil start-address halt-address
         nil nil nil 0 *popcount-64* x86))
       \textitg{;; Input n is symbolic and located in rdi.}
       (x86 (wr64 *rdi* n x86))
       \textitg{;; 300 is the upper bound on the number of steps to take.}
       \textitg{;; Execution halts if the halt-address is reached earlier.}
       (x86 (x86-run 300 x86)))
    (and
     \textitg{;; No error was encountered during state initialization.}
     (equal flg nil)
     \textitg{;; rax contains the popcount of n.}
     (equal (rgfi *rax* x86) (logcount n))
     \textitg{;; rip contains the address of the instruction following the}
     \textitg{;; halt address.}
     (equal (rip x86) (+ 1 halt-address))))
  :g-bindings
  `((n   (:g-number ,(gl-int 0 1 65)))))
\end{Verbatim}

Of course, such a theorem is not useful if re-compiling the same
high-level program results in a different machine-code program, which
may be located at a different linear memory location or may use
different x86 instructions altogether.  The former case would not have
been a problem if {\tt x86-popcount-64-correct} was a statement about
the position-independent version of the program, whereas the latter
case is a downside of machine-code verification in general.  The
benefit of this approach is that it provides a simple solution in both
these situations.  A user can simply re-submit the new conjecture to
ACL2 so that the proof proceeds completely automatically as before.

GL may reach its limits if used to verify the second kind of program,
which is structurally complex and whose proof involves determining
inductive invariants.  One can reason about this program on the x86
ISA model using the classical {\em clock functions}
approach~\cite{ray-mechanical}.  A clock function computes the number
of steps needed for the program to reach a desired state.  The
program's correctness can be stated as follows: given a state {\tt
  x86$_\mathrm{i}$} that satisfies some preconditions, the final state
{\tt x86$_\mathrm{f}$ = x86-run(n, x86$_\mathrm{i}$)}, where {\tt n}
denotes the (possibly symbolic) value computed by the clock function,
satisfies some postconditions.  For instance, we verified a word-count
program that makes system calls to read input from a file --- the
\href{https://github.com/acl2/acl2/tree/master/books/projects/x86isa/proofs/wordCount/wc.lisp}{\underline{proof
    script}} and a detailed description~\cite{Goel:Syscalls} are
available.  One of the final theorems of correctness for this program
is as follows.  The function {\tt preconditions} specifies general
conditions for the correctness of this program --- e.g., the program
is located at a suitable symbolic address {\tt addr} in the initial
x86 state, whose various components contain symbolic values.
\begin{Verbatim}[baselinestretch=1,commandchars=\\\{\}]
(defthm program-behavior-nc
  (implies
   (and (preconditions addr x86)
        (equal offset (offset x86))
        (equal str-bytes (input x86)))
   (equal
    \textitg{;; nc: gets the number of characters computed by the program}
    \textitg{;; from the x86 state}
    (nc (x86-run (clock str-bytes x86) x86))
    \textitg{;; nc-spec: specification function that computes the number}
    \textitg{;; of characters}
    (nc-spec offset str-bytes))))
\end{Verbatim}
Though the amount of user interaction required is higher in this case,
one obtains a more general correctness theorem than that for the first
program.  The applicability of this theorem is still adversely
affected in case the machine-code program generated by re-compiling
the source program contains different x86 instructions.  However, it
is possible that one may be able to re-use some lemmas from the proof
script.

In addition to functional correctness, the {\tt x86isa} books provide
lemmas that help in the proof of other kinds of properties.  For
instance, for the word-count program, we proved that the values in all
memory locations, except the program's stack, in the final x86 state
are exactly the same as that in the initial state.  In other words,
this theorem states that at the end of its execution, the word-count
program did not change any values in unintended regions of memory.

We note that the position-independent version of the popcount program
can also be verified using the clock functions approach without
incurring too much overhead.  The lemmas supporting symbolic
simulation in the {\tt x86isa} books can help in automatically
obtaining the ACL2 expression corresponding to the value in the {\tt
  rax} register in the final x86 state, and GL can be used to prove
that this expression computes the same value as that computed by {\tt
  logcount}.  This ACL2 expression obtained after symbolic simulation
will change if the instructions in the machine program change due to
re-compilation of the source code.  In this case too, one can use GL
to automatically prove the equality of this expression with {\tt
  logcount}.  Thus, not only does this approach win us a general
theorem of correctness, but it also provides a relatively cheap way of
re-proving a general correctness theorem in case the program changes.
In this manner, GL can be used to reason about computationally
intensive pieces of code in a structurally complex program, and lemmas
provided by the {\tt x86isa} books can be used to perform
compositional reasoning to obtain the proof of correctness of the
entire program.  More details can be found along with the popcount
\href{https://github.com/acl2/acl2/tree/master/books/projects/x86isa/proofs/popcount/popcount-general.lisp}{\underline{proof
    script}}.

\subsection{Verifying System Programs}\label{sec:system-prog-verification}
System programs can also be verified using the same strategies as
application programs.  However, generally speaking, the reasoning
overhead of system programs is higher because they access and modify a
larger x86 state than application programs.  In this paper, we focus
only on the upshot of the processor's side-effect updates to {\em
  accessed} and {\em dirty flags} during address translation via
paging.

The accessed and dirty flags are two fields present in the entries of
the paging data structures.  Whenever an entry is referenced during an
address translation, the processor sets its accessed flag.  When the
translation is done on behalf of a memory write operation, then the
processor sets the dirty flag in the final entry that points to the
physical address.  Effectively, these updates are side-effects of the
processor as it works to translate a linear address.  Thus, all linear
memory operations --- including memory reads --- modify the memory, as
a result of which one needs to establish non-interference (or overlap)
properties about every linear memory operation.  These side-effect
updates are numerous --- merely fetching one byte of an x86
instruction from the memory can cause many side-effect updates.  The
sheer number of these side-effect updates increases reasoning overhead
significantly.

The system-level mode of operation offers two sub-modes --- marking
and non-marking mode --- that are exactly the same except in their
treatment of these side-effect updates.  The marking mode of operation
specifies these side-effect updates to the accessed and dirty flags,
whereas the non-marking mode suppresses them.  For supervisor-mode
programs that do not depend on these side-effect updates, we recommend
verifying the program in the non-marking mode and then porting the
proof over to the marking mode.  This is because of the simpler linear
memory non-interference theorems in the non-marking mode --- these
theorems have fewer hypotheses in the non-marking mode because one
does not have to preclude reads from those regions of the memory that
are changed by the side-effect updates.  Porting the proof over to the
marking mode is simply a matter of adding relevant (and mostly
obvious) disjointness preconditions to the theorems --- for example,
the paging entries that govern the translation of the program and the
program itself must be disjoint.  Note that this condition was
unnecessary in the non-marking mode because the paging entries of the
program are not modified over the course of its execution. The {\tt
  x86isa} library contains books that includes a proof of correctness
of a supervisor-mode
\href{https://github.com/acl2/acl2/tree/master/books/projects/x86isa/proofs/zeroCopy}{\underline{zero-copy}}
program that manipulates the paging data structures; this proof
illustrates this methodology of first using the non-marking mode and
then the marking mode to verify system programs.

%% file: future-plans.tex
\section{How Do I...?}
We anticipate and answer some specific how-to questions that may be
asked by a user and potential future contributor to the {\tt x86isa}
books.\newline

\noindent {\bf How do I add a new component to the x86 state?}\newline
The x86 state is defined using an abstract
stobj~\cite{abstract-stobjs} layered on top of a concrete
stobj~\cite{stobjs}.  The ``default'' abstract representation in our
model for simple fields is the same as the logical representation of
the concrete field; for array fields, it is a record~\cite{records}
with a default value of 0.

Suppose one wanted to add the (currently missing) 64-bit Extended
Control Register {\tt xcr0} to the x86 state.  One must first add a
suitable field to the concrete stobj {\tt x86\$c}.  Note that {\tt
  xcr0\$c} is a simple (non-array) field.
\begin{Verbatim}
(xcr0$c :type (unsigned-byte 64) :initially 0)
\end{Verbatim}
The {\tt x86isa} books contain macros that use the above expression to
generate suitable events that will help in defining and admitting the
corresponding abstract stobj.  For instance, the abstract and
top-level recognizer, accessor, and updater functions, along with the
correspondence function pertaining to this field will be automatically
generated.  One would have to resolve the proof obligations that
establish the correspondence between the concrete and abstract stobjs
(these obligations are generated automatically by the {\tt
  defabsstobj} event), but these will be straightforward for
``default'' abstract representations.

If a different abstract representation for the new component is
required, one would have to disable the automated generation of events
for this component and manually define the appropriate events.  The
memory model in our x86 state is an example of where we used this
manual approach --- the correspondence relation between the concrete
and abstract representation of the x86 memory is complicated.  Memory
is implemented using accessor, updater, and recognizer functions
operating on three distinct fields in the concrete stobj, and these
functions have been proved to correspond to those operating on a
single record field in the abstract stobj.
See~\cite{abstract-stobjs,hunt-kaufmann-fmcad-2012} for details.

We now discuss a modeling quirk of MSRs (machine-specific registers),
since it is likely that a user might want to add more MSRs than are
currently supported by our x86 model.  The x86 ISA defines several
architecture-specific MSRs (possibly, hundreds), but we model only six
of them using an array field in the concrete stobj and a corresponding
record in the abstract stobj.  In order to add a new MSR, simply
increase the number of elements of the {\tt msr\$c} field.  The caveat
here is that unlike other registers, the index of an MSR in our x86
state does not match its identifier on the real machine.  For example,
the $0^{th}$ general-purpose register in the x86 state of our model is
the {\tt rax}, and $0$ is also its identifier on the real machine.  It
suffices to define one constant --- {\tt *rax* = 0} --- pertaining to
this register.  However, the $0^{th}$ MSR in our x86 state corresponds
to the {\tt ia32\_efer} register, whose identifier on the real machine
is {\tt 0xC000\_0080}.  Thus, we define two constants pertaining to
MSRs: one for the real identifier and one for indexing into the {\tt
  msr} field of our model.  For instance, {\tt *ia32\_efer*} is equal
to {\tt 0xC000\_0080} whereas {\tt *ia32\_efer-idx*} is equal to $0$.
If an x86 instruction uses an identifier equal to {\tt *ia32\_efer*},
our specifications use {\tt *ia32\_efer-idx*} to access this
register.\newline

\noindent {\bf How do I add a new x86 instruction?} \newline
There are four main tasks here:
\begin{enumerate}

\item {\em Add the capability to decode the instruction:} The Intel
  manuals~\cite{intel-manuals} have various tables that contain the
  decoding information of x86 instructions, such as their addressing
  information (e.g., whether the instruction uses a {\tt ModR/M} byte
  to determine the location of its operands), default sizes, etc.  The
  {\tt x86isa} books have the Lisp/ACL2 representation of these tables
  (see
  \href{https://github.com/acl2/acl2/tree/master/books/projects/x86isa/utils/decoding-utilities.lisp}{\underline{this
      book}}) --- any instruction with a one- or two-byte x86 opcode
  can be decoded using our ACL2 functions that read information off
  these tables.  For all other instructions (e.g., those that have
  three-byte opcodes), one would have to manually port some more
  relevant tables from the Intel manuals into the {\tt x86isa} books.

\item {\em Write the instruction semantic function:} The {\tt x86isa}
  books supply a {\tt def-inst} macro to specify instruction semantic
  functions.  This macro is simply a wrapper around
  \href{http://www.cs.utexas.edu/users/moore/acl2/manuals/current/manual/?topic=ACL2____DEFINE}{{\tt
      \underline{define}}} --- it also adds the instruction's details
  to a table automatically so that one can keep track of the opcodes
  supported by the {\tt x86isa} books.

\item {\em Extend the step function:} Depending on the opcode of an
  instruction, the step function {\tt x86-fetch-decode-execute}
  dispatches control to an appropriate instruction semantic function.
  Simply invoke the new instruction semantic function from the step
  function.

\item {\em Validate the instruction's specification:} Using utilities
  described in Section~\ref{sec:concrete-program-run}, run
  co-simulations of the model against the real machine to validate the
  new instruction semantic function.\newline

\end{enumerate}

\noindent {\bf How do I abstract away the behavior of a standard C
  library function, say {\tt printf} or {\tt scanf}?}\newline
As described earlier, the user-level mode of operation extends the
semantics of the {\tt syscall} instruction to support both reasoning
and execution of some OS-provided system calls.  Standard C library
functions like {\tt printf} and {\tt scanf} are built on top of these
system calls.  One may want to assume the correctness of these library
functions instead of the lower-level system calls when reasoning about
an application program.

One solution is to create yet another mode of operation of the x86 ISA
model --- say, a {\em strong} user-level mode --- where the semantics
of {\tt call}, {\tt jmp}, and any other branch/control-flow
instruction have been extended to support these standard library
functions.  Note that similar to system calls, these functions will be
non-deterministic from the point of view of the application.  Thus,
one can use the {\tt env} field to model the external environment that
these library functions depend on.

\section{Potential Future Projects}

Though the {\tt x86isa} books model a significant portion of the
x86-64 ISA, they are incomplete.  Also, there are several ways in
which its lemma libraries can be improved.  We now discuss some short-
and long-term projects that, once completed, can improve the quality
and feature-set of {\tt x86isa}.  Of course, this list is
non-exhaustive.

\subsection*{ISA Modeling Projects}

\begin{itemize}
\item {\bf {\em Exceptions and Interrupts:}} As of this writing, the
  {\tt x86isa} books model system registers relevant to both
  exceptions and interrupts ({\tt idtr}, {\tt gdtr}, {\tt ldtr}, and
  {\tt tr}) and contain a specification of the Interrupt, Global, and
  Local Descriptor Tables (IDT/GDT/LDT), including recognizers for
  well-formed table entries or {\em gates}.  These gates contain
  information about the location of the interrupt- or
  exception-handling procedures.  The {\tt x86isa} books also support
  system instructions like {\tt lgdt}, {\tt lldt}, and {\tt lidt} used
  to initialize the system registers.

  We already support exceptions in a limited sense --- whenever we
  detect that an error condition is reached (for example, if a {\tt
    div} instruction's divisor operand is 0, which corresponds to a
  {\tt \#DE} exception; or if a paging entry encountered during a page
  walk is ill-formed, which corresponds to a {\tt \#PF} exception), we
  populate the {\tt ms} field with some information relevant to this
  exception and halt the interpreter.  Currently, the preconditions
  for the correctness of programs analyzed using {\tt x86isa} weed out
  all such erroneous conditions.  In order to support exceptions in
  their entirety, the appropriate exception-handling procedures must
  be called by looking up relevant information in the descriptor
  tables --- the current solution of populating the {\tt ms} is only a
  stopgap.  On the other hand, interrupts are asynchronous events
  (unlike exceptions) and their implementation is likely to require
  some significant changes/additions to the x86 ISA model.  For
  example, since all interrupts are guaranteed to be taken on an
  instruction boundary, we could consult an oracle to check for the
  occurrence of some interrupt at every such boundary.  Upon
  encountering one, we can transfer control to the appropriate
  interrupt-handling procedure.  Note that one would need to modify
  the step and/or the run functions if we adopt this approach.

  Given that much of the support required for implementing both
  exceptions and interrupts already exists in the x86 model, we
  estimate that this will be a relatively short-term project.

\item {\bf {\em I/O Capabilities:}} I/O instructions like {\tt in} and
  {\tt out} are not supported by {\tt x86isa} yet.  Implementing them
  will also be a short-term project because the existing
  infrastructure around {\tt env} can be re-used to characterize
  interaction of the processor with an external environment.

\item {\bf {\em Caches and Multiprocessors:}} Our x86 model can be
  extended to include the entire memory hierarchy --- including
  caches, translation-look aside buffers, store buffers, etc. --- in
  order to obtain a complete specification of how memory reads and
  writes are resolved by multiple processors.  This promises to be
  quite a long-term project because it would involve dealing with
  complicated properties like cache coherence.

\end{itemize}

We briefly comment on a challenge that a contributor to the {\tt
  x86isa} books is likely to face when modeling some advanced features
of the x86 ISA, such as interrupts or protection management.  Ideally,
to validate the x86 ISA model, one must co-simulate it against a
``bare'' x86 processor, i.e., one not running any OS.  However, a bare
x86 machine is difficult to work with, and so far we have validated
our model against an x86 ISA processor running a mainstream OS like
Linux\footnote{Of course, one could also choose to validate the x86
  ISA model against a hardened instruction-set simulator like QEMU.}.
Unfortunately, an OS is tightly inter-twined with the workings of a
processor's system features, thereby making it difficult to separate
OS-specific behavior from the machine's behavior.  We posit that a
satisfactory way to validate the specification of system features is
by running a mainstream (if stripped down) OS on the x86 ISA model.
This task will require adding several features and instructions
currently missing from {\tt x86isa}.  Not only will this undertaking
increase confidence in the accuracy of the x86 model, but it will also
enable us to reason about {\em real} system code that is deployed on
modern machines.  Needless to say, this is a formidable long-term
project, but one with high returns.

\subsection*{Proof-related Projects}

\begin{itemize}

\item {\bf {\em Using Codewalker:}} One can imagine using
  \href{https://github.com/acl2/acl2/tree/master/books/projects/codewalker}{\underline{Codewalker}}
  to lift reasoning about x86 machine code to high-level specification
  functions.  This project may involve adding capabilities to the
  Codewalker and/or {\tt x86isa} books.

\item {\bf {\em Automated Precondition Discovery:}} A challenging
  aspect of program verification is discovering the preconditions
  under which a program behaves as expected.  One way in which ACL2
  users figure out some of the preconditions is by observing the
  reason why some rules fail to fire when expected, and then adding
  any missing hypotheses to the conjecture that make those rules
  applicable to the goal at hand.  A useful project could be to
  automate this process so that after a failed proof attempt, a user
  is presented with hypotheses that are good candidates to be
  top-level preconditions.  One can imagine such a capability to be
  useful in other ACL2 projects too.  As such, this project can easily
  be a long-term one.

\end{itemize}

%% file: conclusion.tex
\section{Conclusion}\label{sec:conclusion}
This paper serves as a good starting point for a user or a potential
future developer of the {\tt x86isa} books.  We recommend that a new
user and/or someone with little experience in low-level code
verification begin by executing some concrete runs of a program on the
x86 model before moving on to program verification.  Also, it is
advisable --- for both reasoning and execution --- to first use the
x86 model in its user-level mode instead of the more complicated
system-level mode of operation.

We give an overview of the development style and capabilities of the
{\tt x86isa} books in this paper.  However, a more complete
description is available~\cite{goel-dissertation} --- this work's
accompanying Ph.D. dissertation describes how the x86 model is
optimized for both reasoning and execution efficiency, how complicated
x86 ISA mechanisms such as IA-32e paging and segmentation are
specified, how congruence-based rewriting is used to reduce reasoning
overhead in the system-level mode of operation, and other pertinent
details.  Also, the most up-to-date information about these books is
available at
\href{http://www.cs.utexas.edu/users/moore/acl2/manuals/current/manual/?topic=ACL2____X86ISA}{{\tt
    :doc \underline{x86isa}}}.

There are many ways in which these books can be used and/or extended,
beyond what we discussed in the previous section.  An example of one
such application of this work, not related to program verification, is
micro-architecture verification --- e.g., one can use the x86 ISA
model to prove that one or more x86 micro-operations implement an
ISA-level instruction.  This work paves the way for research and
engineering opportunities that would otherwise have been difficult to
pursue --- we hope that the community gets involved in this project.